\documentclass{jetpl}
\usepackage{graphicx}
\usepackage{amssymb}
\usepackage{amsmath}
\usepackage{color}
\usepackage{hyperref}
\twocolumn
\newcommand*\Bell{\ensuremath{\boldsymbol\ell}}

\title{Superfluid $^3$He in squeezed nematic aerogel}
\rtitle{Superfluid $^3$He in squeezed nematic aerogel}
\sodtitle{Superfluid $^3$He in squeezed nematic aerogel}

\author{V.\,V.\,Dmitriev$^+$\thanks{e-mail: dmitriev@kapitza.ras.ru},\,M.\,S.\,Kutuzov$^*$,\,A.\,A.\,Soldatov$^+$,\,A.\,N.\,Yudin$^{+,\circ}$}
\rauthor{V.\,V.\,Dmitriev,\,M.\,S.\,Kutuzov,\,A.\,A.\,Soldatov,\,A.\,N.\,Yudin}
\sodauthor{V.\,V.\,Dmitriev,\,M.\,S.\,Kutuzov,\,A.\,A.\,Soldatov,\,A.\,N.\,Yudin}

\address{$^+$P.\,L. Kapitza Institute for Physical Problems of RAS, 119334 Moscow, Russia}
\address{$^*$Metallurg Engineering Ltd., 11415 Tallinn, Estonia}
\address{$^\circ$National Research University Higher School of Economics, 101000 Moscow, Russia}

\dates{\today}{*}

\begin{document}

\abstract{We present results of nuclear magnetic resonance (NMR) experiments in superfluid $^3$He in two samples of nematic aerogel consisting of nearly parallel mullite strands. The samples were cut from the same piece of the aerogel, but one of them was squeezed by 30\% in the direction transverse to the strands. In both samples the superfluid transition of $^3$He occurred into the polar phase, where no qualitative difference between NMR properties of $^3$He in these samples was found. The difference, however, has appeared on further cooling, after the transition to the polar-distorted A phase (PdA phase) with the orbital part of the order parameter in the 2D Larkin-Imry-Ma (LIM) state. In the squeezed sample the 2D LIM state is anisotropic that results in changes in the NMR, which can be used as an additional marker of the PdA phase and have allowed us to measure the value of the anisotropy.}

\maketitle

\section{Introduction}
Nematic aerogels consist of strands with diameter of $\sim10$\,nm, which are oriented along the same direction \cite{asad15}.
In liquid $^3$He in these aerogels, the strands result in an anisotropy of $^3$He quasiparticle scattering \cite{meln15}. It makes favorable new superfluid phases: polar, polar-distorted A (PdA) and polar-distored B (PdB) \cite{AI,fom14,Ik}. The superfluid order parameter of polar and PdA phases has the form:
\begin{equation}\label{par}
A_{\nu k}=\Delta_0 e^{i\varphi}d_\nu(am_k+ibn_k),
\end{equation}
where $\Delta_0$ is the gap parameter, $e^{i\varphi}$ is the phase factor, $\bf d$ is the unit spin vector oriented normal to the magnetization, $\bf m$ and $\bf n$ are mutually orthogonal unit vectors in orbital space, and $a^2+b^2=1$. The PdA phase ($a^2>b^2$) is an intermediate state between the polar phase ($a=1, b=0$) and the A phase ($a=b$) which exists in bulk $^3$He. The A and PdA phases are chiral and have a superfluid gap with two nodes along $\Bell={\bf m}\times{\bf n}$ while the polar phase is not chiral, it is characterized only by one orbital vector $\bf m$, and its gap is zero in the plane normal to $\bf m$. The phases described by Eq.\eqref{par} belong to a class of Equal Spin Pairing (ESP) states.

Previous experiments with $^3$He in nematic aerogel were done using samples with strands of AlOOH (Obninsk aerogel) or Al$_2$O$_3$ (nafen) with various porosities and degree of the anisotropy \cite{dmit12,sur14,dmit15,ask15,parp,fin,dmit18}. It was found that on cooling from the normal phase, the superfluid transition of $^3$He occurs into the PdA phase (in Obninsk aerogel) or into the polar phase (in nafen). On further cooling, transitions from polar to PdA phase, and then into PdB phase were observed.

Here, using continuous wave (cw) nuclear magnetic resonance (NMR), we investigate the ESP phases of $^3$He in a new type of nematic aerogel, strands of which are made of mullite. We assume that the mullite aerogel is closer to an ideal array of parallel cylinders because it is more transparent than Obninsk aerogel or nafen and more easily splits along the strands. Two samples of mullite nematic aerogel are used: the original (undeformed) sample and the sample squeezed by 30\% in the direction transverse to the strands. In particular, we investigate how the squeezing changes NMR properties of chiral PdA phase and non-chiral polar phase.

\section{Samples and methods}
Two samples were cut from the same piece of aerogel with overall density 150\,mg/cm$^3$, porosity $\approx96$\%, and diameter of strands $\lesssim10$\,nm. One sample (mullite-F) was placed freely in a separate cell of our experimental chamber with a filling factor $\approx80$\%, another sample (mullite-S) was unidirectionally squeezed by 30\% inside its cell in the direction transverse to the strands by a movable wall that was glued after. Both samples have a cuboid shape of sizes 3--4\,mm. The chamber was made of Stycast-1266 epoxy and was similar to that described in Ref.~\cite{dmit12}.

Measurements of spin diffusion in normal $^3$He in present samples were done at 2.9\,bar using spin echo techniques. It allowed us to determine effective mean free paths of $^3$He quasiparticles along ($\lambda_\parallel$) and transverse to the strands ($\lambda_\perp$) in the limit of zero temperature (see Table~\ref{table1}).

\begin{table}[t]
\center
\caption{Table\,\thetable.
Characteristics of the samples of different nematic aerogels: nafen-90 and nafen-243 (used in previous experiments, the data are taken from Ref.~\cite{meln15}), mullite-F and mullite-S (used in present experiments)}
\label{table1}
\begin{tabular}{c|c|c|c}
\hline
Sample & Porosity (\%) & $\lambda_\parallel$ (nm) & $\lambda_\perp$ (nm) \\
\hline
nafen-90 & 97.8 & 960 & 290 \\
nafen-243 & 93.9 & 570 & 70 \\
mullite-F & 96 & 900 & 235 \\
mullite-S & 94.3 & 550 & 130 \\
\hline
\end{tabular}
\end{table}

Experiments were carried out using linear cw NMR in magnetic fields 139--305\,Oe (NMR frequencies 450--990\,kHz) at pressures 7.1--29.3\,bar. We were able to rotate the external steady magnetic field $\bf H$ by an arbitrary angle $\mu$ with respect to the direction of strands ($\hat z$) as shown in Fig.~\ref{scheme}a. Additional gradient coils were used to compensate the magnetic field inhomogeneity. The necessary temperatures were obtained by a nuclear demagnetization cryostat and determined using cw NMR signal from bulk A phase (in gaps between the sample and cell walls) or by a quartz tuning fork calibrated by measurements of the Leggett frequency in bulk $^3$He-B. To avoid a paramagnetic signal from surface solid $^3$He and stabilize the polar phase \cite{dmit18}, the samples were preplated by $\sim 2.5$ atomic layers of $^4$He.

\section{Theory}
\begin{figure}[t]
\center
\includegraphics[width=1\columnwidth]{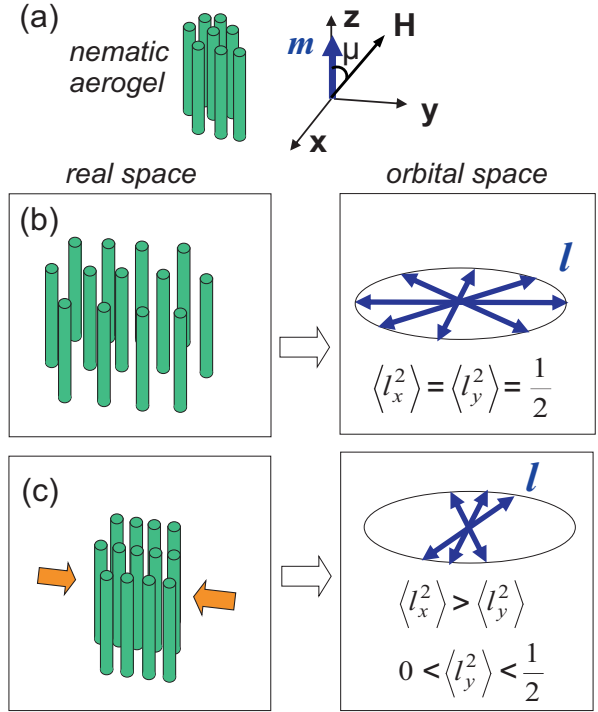}
\caption{Fig.\,\thefigure.
The orientation of $\bf m$ in the ESP phases and $\bf H$ with respect to the strands of nematic aerogel (a) and the distribution of $\Bell$ in PdA (or A) phase in the original sample (b) and the sample squeezed perpendicular to the strands (c)}
\label{scheme}
\end{figure}

In the ESP phases described by Eq.\eqref{par} strands of nematic aerogel fix ${\bf m}\parallel \hat z$ (Fig.~\ref{scheme}a) and destroy the long-range order in the A and PdA phases, where $\Bell$ forms a static two-dimensional Larkin-Imry-Ma (2D LIM) state \cite{ask15,vol08}. In this state $\Bell$ is inhomogeneous at distances $\gtrsim1$\,$\mu$m and randomly oriented in the plane normal to the strands. In the undeformed sample the 2D LIM state should be isotropic in this plane; that is, averaged over space, projections of $\Bell$ are: $\left<\ell_x^2\right>=\left<\ell_y^2\right>=1/2, \left<\ell_z^2\right>=0$ (Fig.~\ref{scheme}b). However, in the PdA phase it is possible to create an anisotropic 2D LIM state by squeezing the sample perpendicular to the strands (in our experiments along $\hat y$). Such deformation changes correlations in a spatial distribution of strands \cite{Mel} that orients $\Bell$, on average, along $\hat x$; that is, $\left<\ell_x^2\right>>1/2, \left<\ell_y^2\right><1/2$ (Fig.~\ref{scheme}c), and qualitatively changes NMR properties \cite{ask15}. In non-chiral polar phase the squeezing should not change NMR properties qualitatively.
We note that the vector $\bf d$ in low excitation NMR experiments is spatially homogeneous \cite{10}, and here we consider only this case.

Identification of the ESP phases can be made by measurements of the cw NMR frequency shift ($\Delta\omega$) from the Larmor value ($\omega_L=\gamma H$, where $\gamma$ is the gyromagnetic ratio of $^3$He). The shift appears due to a dipole interaction and depends on the order parameter, its spatial distribution, angle $\mu$, and in the isotropic 2D LIM state it is given by \cite{dmit12,ask15,Min}:
\begin{equation}\label{dwISO}
2\omega_L\Delta\omega=k(4-6b^2)\Omega_A^2\cos^2\mu=K\Omega_A^2\cos^2\mu,
\end{equation}
where $K=k(4-6b^2)$ and, in a weak coupling approximation,
\begin{equation}\label{K}
k=\frac{1}{3-4a^2b^2}.
\end{equation}
Here $\Omega_A$ = $\Omega_A(T,P)\propto \Delta_0$ is the Leggett frequency of the A phase. This frequency grows from 0 up to $\sim$100~kHz on cooling from the superfluid transition.
If the superfluid transition temperature of $^3$He in aerogel ($T_{ca}$) is close to the transition temperature ($T_c$) in bulk $^3$He, then $\Omega_A(T/T_{ca})/\Omega_{A0}(T/T_c)=T_{ca}/T_c$ \cite{dmit15}, where $\Omega_{A0}$ is the Leggett frequency of  bulk $^3$He-A, which is known. Then measurements of $\Delta\omega$ allow to determine $K$ and to identify the phases (in the A phase $K=k=1/2$, while in the polar phase $K$ should be equal to 4/3). However, it is known that the weak coupling limit is a good approximation only at low pressures \cite{VW}, so, for the identification of the phases at high pressures, an experimentally measured dependence of  $K$ in the polar phase ($K_p$) on pressure should be used. Experiments show that $K_p$ decreases from 4/3 at 0 bar down to $\approx1.15$ at 29.3\,bar \cite{dmit15}, so $k$ in the polar phase ($k_p$) decreases from 1/3 down to 0.29.

In squeezed nematic aerogel (Fig.~\ref{scheme}c) Eq.~\eqref{dwISO} is modified for A and PdA phases due to the anisotropy of the 2D LIM state \cite{ask15}. In particular, the shifts $\Delta\omega_{\parallel}$ (at $\mu=0$) and $\Delta\omega_{\perp}$ (at $\mu=\pi/2$ and $\bf H \parallel \hat y$) are given by:
\begin{equation}\label{dwANI0}
2\omega_L\Delta\omega_{\parallel}=4\left(1-b^2-b^2\left<\ell_y^2\right>\right)k\Omega_A^2,~~\mu=0,
\end{equation}
\begin{equation}\label{dwANI90}
2\omega_L\Delta\omega_{\perp}=4b^2\left(1-2\left<\ell_y^2\right>\right)k\Omega_A^2,~~\mu=\pi/2.
\end{equation}
It follows from Eq.~\eqref{dwANI90} that at $\mu=\pi/2$ there is a qualitative difference between PdA (or A) and polar phases: in the polar phase $\Delta\omega_{\perp}=0$, while in the PdA phase the shift is nonzero (positive for $\left<\ell_y^2\right><1/2$). The latter can be used as an additional marker of the PdA phase. At low pressures (where the weak coupling model works) measurements of $\Delta\omega_{\parallel}$ and $\Delta\omega_{\perp}$ allow to determine $b^2$ and the value of anisotropy of the 2D LIM state ($\left<\ell_y^2\right>$) in the PdA phase.
In order to estimate $\left<\ell_y^2\right>$ and $b^2$ at high pressures, we can use experimental value of $K_p=4k_p$ and assume that, for small distortions from the polar state, strong coupling corrections do not change qualitatively the dependence of $k$ on $b^2$; that is, $k(P)/k_p(P)=3/(3-4a^2b^2)$.

\section{Results and discussions}
\begin{figure}[t]
\center
\includegraphics[width=0.95\columnwidth]{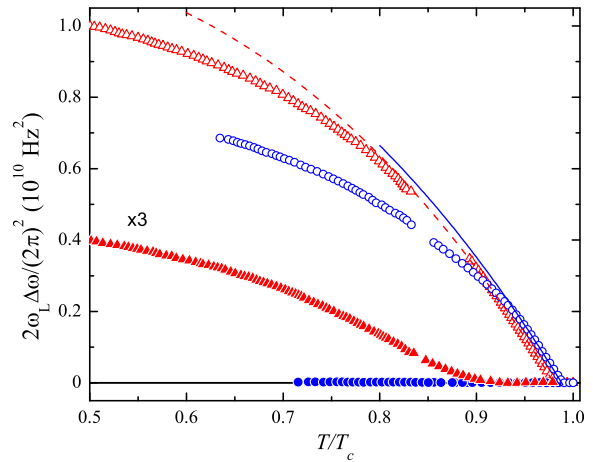}
\caption{Fig.\,\thefigure.
$\Delta\omega_{\parallel}$ (open symbols) and $\Delta\omega_{\perp}$ (filled symbols) versus temperature in mullite-F (circles) and mullite-S (triangles). Solid and dashed lines are the theory for $\Delta\omega_{\parallel}$ in the polar phase with $K_p=1.15$ for $T_{ca}=0.988T_c$ and $T_{ca}=0.980T_c$ respectively. Values of $\Delta\omega_{\perp}$ in mullite-S are multiplied by 3 for better view. $P=29.3$\,bar.}
\label{shifts}
\end{figure}
\begin{figure}[t]
\center
\includegraphics[width=1\columnwidth]{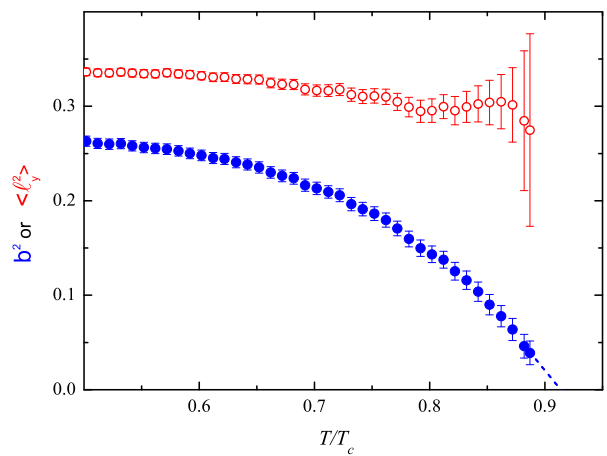}
\caption{Fig.\,\thefigure.
Temperature dependencies of $b^2$ (filled circles) and $\left<\ell_y^2\right>$ (open circles) calculated using Eqs.~\eqref{dwANI0} and \eqref{dwANI90} and the data from Fig.~\ref{shifts} for mullite-S.}
\label{calc}
\end{figure}
\begin{figure}[t]
\center
\includegraphics[width=1\columnwidth]{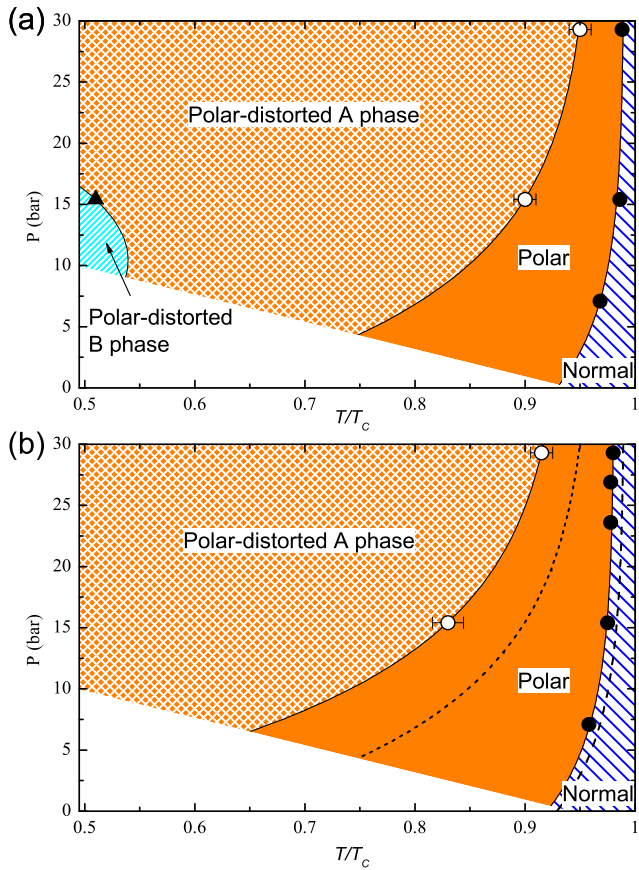}
\caption{Fig.\,\thefigure.
Phase diagrams of $^3$He in mullite-F (a) and mullite-S (b) obtained on cooling from the normal phase. Filled circles mark $T_{ca}$. Open circles mark the transition between the polar and PdA phases. Triangle marks the beginning of the transition into the PdB phase on cooling. The dashed and short dashed lines on the panel (b) indicate the transitions between the normal and polar, the polar and PdA phases respectively for the case of mullite-F. The white area shows regions with no experimental data. The $x$ axis represents the temperature normalized to the superfluid transition temperature of bulk $^3$He.}
\label{pd}
\end{figure}
The superfluid transition of $^3$He in mullite-F occurs into the polar phase at $T=T_{ca}\approx0.988\,T_c$ (at $P=29.3$\,bar) as it is seen from measurements of $\Delta\omega_{\parallel}$ (open circles in Fig.~\ref{shifts}). On further cooling, a second-order transition into the PdA phase takes place at $\approx0.95\,T_c$ as the data deviate from the curve expected for the polar phase. The shift for $\mu=\pi/2$ (filled circles) is absent that indeed agrees with Eq.~\eqref{dwISO}.

In mullite-S the superfluid transition to the polar phase at $P=29.3$\,bar is observed at $T=T_{ca}\approx0.980\,T_c$, and just below this temperature data for $\Delta\omega_{\parallel}$ (open triangles in Fig.~\ref{shifts}) follow the curve with the same slope as for mullite-F. On cooling, the polar phase persists, until the positive shift for $\mu=\pi/2$ (filled triangles) appears at $T=T_{PdA}\approx0.915\,T_c$ indicating a transition to the PdA phase. We note that the data for $\Delta\omega_{\parallel}$ in mullite-S  start clearly deviate from the curve for the polar phase (the dashed line) far below $T_{PdA}$. It agrees with Eq.~\eqref{dwANI0} from which it follows that $\Delta\omega_{\parallel}$ in PdA phase in the anisotropic 2D LIM state is greater than in the isotropic state and can even exceed the shift in polar phase if $4b^2 < 1-3\left<\ell_y^2\right>$.

Using Eqs.~\eqref{dwANI0},\eqref{dwANI90} and the measured (and extrapolated to a given temperature) values of $\Delta\omega_{\parallel}$ and $\Delta\omega_{\perp}$ we have obtained temperature dependencies of $b^2$ and $\left<\ell_y^2\right>$ in the PdA phase (Fig.~\ref{calc}). Points in this figure correspond to $K_p=1.15$ and $T_{ca}=0.980\,T_c$, and error bars show how the results are changed if we vary $K_p$ by 1\%. It is seen that $b^2$ increases from 0 to 0.26 on cooling from $T_{PdA}$ in agreement with Ref.~\cite{AI}, while $\left<\ell_y^2\right>$ levels off at $\approx0.33$ confirming the anisotropy of the 2D LIM state. The same measurements were done at $P=15.4$\,bar, where we have obtained that $\left<\ell_y^2\right>\approx0.35$ and $b^2$ increases from 0 up to 0.13 on cooling from $T=T_{PdA}\approx0.83\,T_c$ down to $0.65\,T_c$.

On cooling to the lowest temperatures in $^3$He in mullite-F we observe a decrease of the spin susceptibility (the cw NMR line intensity) and an abrupt change in $\Delta\omega$ indicating a first-order transition to the non-ESP phase, presumably, the polar-distorted B phase, which was the case in previous experiments with $^3$He in nematic aerogels. Based on our measurements, we obtain superfluid phase diagrams of $^3$He in mullite-F and mullite-S shown in Fig.~\ref{pd}.

In recent theoretical paper \cite{fom18} it was stated that Anderson theorem for s-wave superconductors is applicable to superfluid $^3$He in ideal nematic aerogel; that is, in the case of ideally parallel strands and for specular reflection of $^3$He quasiparticles. In particular, the change of $\Delta\omega_{\parallel}$ near the absolute zero should be proportional to $-T^3$ that was confirmed in recent experiments \cite{Aa}. Our results agree with another prediction of Ref.~\cite{fom18}; that is, the range of temperatures where the polar phase exists is inversely proportional to $\lambda_{\perp}$ (see Fig.~\ref{pd}b and data for $\lambda_{\perp}$ for mullite samples in Table~\ref{table1}). We also note that the suppression of the superfluid transition temperature of $^3$He in mullite-F (with porosity 96\%) is essentially (about 1.4 times) smaller than in $^3$He in nafen-90 with higher porosity (97.8\%) and is nearly the same as in nafen-72 with porosity 98.2\%. It agrees with one more prediction of  Ref.~\cite{fom18} that in the ideal case $T_{ca}\equiv T_c$.

\section{Conclusions}
Using cw NMR we have investigated the ESP phases of $^3$He in two samples of new (mullite) nematic aerogel. In both samples the superfluid transition of $^3$He occurs into the polar phase with no qualitative difference between NMR properties of $^3$He in these samples. But the difference is observed on further cooling, after the transition to the PdA phase with $\Bell$ forming the 2D LIM state, which is anisotropic in the squeezed sample. In the latter case we have determined the value of this anisotropy and the degree of the polar distortion. The results, in general, provide an additional proof of existence of the polar phase at $T_{PdA}<T<T_{ca}$ and also support the application of Anderson theorem to superfluid $^3$He in nematic aerogel.

This work was supported by the Russian Science Foundation (project no. 18-12-00384).

\end{document}